\documentclass[aps,preprint,pra,showpacs]{revtex4}
\usepackage{graphicx}
\usepackage[pdftex]{color}
\def\xt{({\bf x},t)}
\def\x{{\bf x}}
\def\A{{\bf A}}
\def\del{\nabla}
\def\J{{\bf J}}
\def\Lo{{L^z_{e}}}
\def\Lt{{L^z_{c}}}
\def\vo{{\varphi_e}}
\def\vt{{\varphi_c}}
\def\pso{{\Psi_e}}
\def\pst{{\Psi_c}}
\def\half{{1\over 2}}

\begin{document}

\title{Nonlocality of the Aharonov-Bohm Effect}

\author{Yakir Aharonov}
\affiliation{School of Physics and Astronomy, Tel Aviv University, Tel Aviv 6997801, Israel\\ and Schmid College of Science, Chapman University, Orange, CA 92866}

\author{Eliahu Cohen}
\affiliation{H.H. Wills Physics Laboratory, University of Bristol, Tyndall Avenue, Bristol, BS8 1TL, U.K.\\ and School of Physics and Astronomy, Tel Aviv University, Tel Aviv 6997801, Israel}

\author{Daniel Rohrlich}
\affiliation{Department of Physics, Ben-Gurion University of the Negev, Beersheba
8410501 Israel}

\date{\today}

\begin{abstract}
Although the Aharonov-Bohm and related effects are familiar in solid state and high energy physics, the nonlocality of these effects has been questioned. Here we show, for the first time, that the Aharonov-Bohm effect has two very different aspects. One aspect is instantaneous and nonlocal; the other aspect, which depends on entanglement, unfolds continuously over time. While local, gauge-invariant variables may occasionally suffice for explaining the continuous aspect, we argue that they cannot explain the instantaneous aspect.  Thus the Aharonov-Bohm effect is, in general, nonlocal.
\end{abstract}


\maketitle

\section{Introduction}
\label{SIntro}

Topological quantum effects appear in very diverse areas of physics \cite{Lau,Has,Wit,Lus,Kib}.
Insofar as these effects are due to potentials that exert no force on particles passing through them, their historical context is the Aharonov-Bohm (AB) effect \cite{ab}. The conventional statement of the AB effect is that, while electromagnetic scalar $V\xt$ and vector ${\bf A}\xt$ potentials are mere calculational aids in classical mechanics, in quantum mechanics they are an essential part of the formalism:  a charged quantum particle can respond to electromagnetic potentials, without ever passing through an electromagnetic field.  At the same time, only gauge-invariant quantities are measurable, and quantum mechanics is manifestly gauge-invariant. It is, therefore, natural to try to dispense with electromagnetic potentials. Yet attempts to do so, over the years, have been unsuccessful.

In 1927, E. Madelung \cite{mad} rewrote the Schr\"odinger equation as two equations in which only gauge-invariant expressions appear, namely, the probability density $\rho \xt \equiv |\Psi \xt |^2$ and probability current $\J\xt$:
\begin{equation}
\J\xt \equiv
{1\over {2m}} \Psi^* (-i\hbar \del -{e\over c}\A )\Psi
+{1\over {2m}} \Psi (i\hbar \del -{e\over c}\A )\Psi^*~~~~.
\label{current}
\end{equation}
But consider an initial wave function $\Psi (\x ,0)$ that vanishes for all $\x$ outside two disjoint regions of radius $a$, e.g. an initial wave function superposing two non-overlapping wave packets, each with the form
\begin{equation}
\psi (\x) =\cases {e^{-1/(a^2 -\x^2 )}  &if $\vert \x\vert \le a$~~~,\cr
                   0                    &if $\vert \x\vert \ge a$~~~~.\cr}
\label{disjoint}
\end{equation}
Madelung's gauge-invariant equations, together with the initial probability density $\rho (\x ,0)$ and probability current $\J(\x ,0)$, do not determine $\Psi \xt$ for all times $t>0$.  For example, suppose we multiply one wave packet by a phase factor, changing the relative phase of the wave packets.  Evidently, as the wave packets evolve and overlap, the resulting interference pattern will depend on the relative phase.  However, neither $\rho (\x ,0)$ nor $\J(\x ,0)$ can, by definition, depend on the relative phase, so Madelung's rewriting of the Schr\"odinger equation was incomplete.

An apparent objection to this argument is that it is physically implausible:  there is no such thing as an infinite potential, and therefore Eq. (\ref{disjoint}) does not correspond to any realistic wave function; $\psi (\x)$ cannot vanish identically in any region.  Hence the wave packets must overlap, and the overlap---even of their exponentially small tails---defines their relative phase.  However, this objection is itself physically implausible, because it makes the interference pattern that evolves from the two wave packets depend in a singular way on the electron's wave function in a region where the probability of finding the electron is exponentially small.  Indeed, the initial phase between the wave packets does not itself determine the AB effect.

Recently, Vaidman \cite{lev} proposed an explanation for the AB effect, via (local) forces rather than via electromagnetic potentials. (See \cite{kk} for an independent but related proposal.) For the magnetic effect, he considers a ``solenoid" made of two counter-rotating, oppositely charged coaxial cylinders. (A solenoid is a coil, but we adopt his terminology.)  He notes that even if the magnetic field of the solenoid is screened from the electron diffracting around it, the transient magnetic field of the passing electron, which is not screened from the rotating cylinders, either increases or decreases the relative rotation rate of the cylinders, according to whether the electron passed on one side or the other of the solenoid.  Entanglement thus develops between the electron and cylinders. The overall wave function of the electron and solenoid is a superposition of two terms, one for each electron path (with corresponding solenoid motion); and their relative phase---the AB phase---results from the torques induced by the transient magnetic field of the electron.

Are the potentials, then, dispensable?  In a previous work \cite{acr} we answered this question in the negative, discussing cases that require a description in terms of potentials. Here, we underline the nonlocality of topological quantum effects by showing that the AB effect has two very different aspects. One aspect is instantaneous and nonlocal; the other aspect, which depends on entanglement, unfolds continuously over time. Although local, gauge-invariant variables may in some cases suffice for explaining the continuous aspect (see also \cite{Berry}), they provide no explanation of the instantaneous aspect. We also challenge, in general, the analysis of the continuous effect via local forces. For example, if a superconducting shield surrounds the flux (see next section), there is no measurable force. Ref. \cite{acr} contains other examples.

\section{A simple model for the vector Aharonov-Bohm effect}
\label{S1}

Let us consider a simple model for the vector (magnetic) AB effect.  In this model, the source of magnetic flux is a long, uniformly charged cylinder of radius $R$ rotating around the $z$ axis. Along the $z$ axis runs a uniform, oppositely charged wire, such that the electric field outside the cylinder vanishes.  Let an electron encircle the cylinder at a constant distance from the $z$ axis.  Denote the moments of inertia of the cylinder and electron as $I_c$ and $I_e$ respectively, and their angular displacements as $\vt$ and $\vo$, respectively.  An appropriate Lagrangian for the cylinder and electron is
\begin{equation}
\mathcal{L} = \half I_c{\dot \vt}^2+ \half I_e {\dot \vo}^2+ I_c \lambda {\dot \vt}{\dot \vo} ~~~,
\label{L}
\end{equation}
where the (dimensionless) coupling $\lambda$ is inversely proportional to the mass of the cylinder.  The $\lambda$ term couples the (angular) speed of the electron ${\dot \vo}$ to $R^2 {\dot \vt}$, which is proportional to the magnetic flux inside the cylinder and to the corresponding vector potential outside it (in a gauge with symmetry about the $z$ axis).  It is convenient to rewrite $\mathcal{L}$ in a simpler form
\begin{equation}
\mathcal{L} = \half I_c\left( {\dot \vt} +\lambda {\dot \vo} \right)^2+ \half I_e^\prime {\dot \vo}^2
\label{LL}
\end{equation}
by renormalizing $I_e$ to $I_e^\prime = I_e\left( 1-{{I_c \lambda^2}/ I_e}\right)$; and for small $\lambda$, it is convenient to substitute $I_e$ for $I_e^\prime$.
Defining the conjugate (angular) momenta $\Lt$ and $\Lo$,
\begin{eqnarray}
\Lt &=& \delta \mathcal{L} /\delta \vt =I_c ( {\dot \vt} +\lambda {\dot \vo})~~~,\cr
\Lo &=& \delta \mathcal{L} /\delta \vo =\lambda I_c ( {\dot \vt} +\lambda {\dot \vo}) +I_e {\dot \vo} ~~~,
\end{eqnarray}
we obtain the (quantum) Hamiltonian as
\begin{equation}
{H} = \Lt {\dot \vt} +\Lo{\dot \vo} - \mathcal{L} ={{(\Lt)^2} \over{2I_c}} +
{{(\Lo -\lambda \Lt)^2} \over{2I_e}}~~~~.
\end{equation}
From $H$ we infer that $\Lt$ and $\Lo$ are constants of the motion. The eigenvalues of $H$ are therefore
\begin{equation} \label{energylevels}
E_{mn}=\frac{\hbar^2}{2}\left[\frac{n^2}{I_c}+\frac{(m-\lambda n)^2}{I_e}\right]~~~~.
\end{equation}

For a well-defined AB phase, the flux through the cylinder, and therefore the angular velocity $\Lt/I_c$, must be well defined.  So let us assume that the cylinder is prepared in an eigenstate of $\Lt$ with eigenvalue $n\hbar$, namely $\pst (\vt)= e^{in \vt}/\sqrt{2\pi}$.  The electron, however, is localized to some $\varphi_0$, as follows \cite{CVA}:
\begin{eqnarray}\label{expandit}
\pso (\vo)&=&{1\over 2\pi}\sum_{m=-\infty}^\infty e^{-m^2/(\Delta m)^2}e^{im(\vo-\varphi_0)}\cr
&=&{1\over 2\pi}\sum_{m=-\infty}^\infty e^{-\left[ {m/ {\Delta m}}-i{{(\vo-\varphi_0)\Delta m}/ 2}  \right]^2}e^{-(\vo-\varphi_0)^2 (\Delta m)^2/4}~~~~.
\end{eqnarray}
Equation (\ref{expandit}) shows that when $\Delta m$ is large, $\pso(\vo)$ is a coherent function of angular displacement, i.e. $\vo \approx \varphi_0$.  (For $\Delta m \rightarrow \infty$, Eq. (\ref{expandit}) reduces to a delta-function in $\vo-\varphi_0$.)

Note that, for $I_c \gg I_e$, $\lambda \vo + \vt$ is a constant of the motion (commutes with the second term in $H$); then every shift $\delta \vo$ in the electron angle $\vo$ induces a corresponding displacement $-\lambda \delta \vo$ in $\vt$.
The unitary operator $e^{i(\lambda \Lt-\Lo)\delta\vo/\hbar}$ commutes with $H$ at all times; applied to the combined wave function $\pso(\vo)\pst(\vt)$, it yields
\begin{equation}
\pso(\vo)\pst(\vt) \rightarrow e^{i\lambda\Lt \delta\vo/
\hbar} \pso(\vo -\delta\vo) \pst(\vt)~~~~.
\end{equation}
We see that $\vo$ shifts by $\delta \vo$; but since $\pst$ is an eigenfunction of $\Lt$, the corresponding ``shift" in $\vt$ is a phase, with $\Lt$ replaced by its eigenvalue $n\hbar$.
The overall wave function acquires a phase $\lambda n\delta\vo$. Thus, as one may also conclude from the semi-classical analysis of Vaidman \cite{lev}, the AB phase of the electron seems to have a local interpretation via the angular shift of the cylinder. However, as we will see in the next section, this local interpretation implies that the physics of the electron must take into account all other (possibly remote) fluxes, not just those that lie between the electron's paths.

Moreover, let us now express $\Lt$ as $\Lt = \delta \Lt+\langle \Lt\rangle$, where $\langle \Lt\rangle =\langle \pst | \Lt |\pst \rangle$ is the average angular momentum of the cylinder. This shift in $\Lt$ is equivalent to a gauge transformation of $\exp(i\vt \langle \Lt \rangle/\hbar)$ which can be physically achieved by adding a singular flux.  When evaluating $(\Lt)^2 = (\delta \Lt)^2 +2 (\delta \Lt) \langle \Lt \rangle +\langle \Lt\rangle^2$, we see that the first two terms induce a negligible phase in the cylinder while the last term is a constant that does not affect the dynamics; so we are only left with the coupling $\lambda\langle \Lt\rangle \Lo$. Then we can understand the change in phase as due to a (constant) average $\langle \Lt \rangle$ rather than to the angular momentum operator $\Lt$.  (See also the work of Aharonov and Anandan \cite{AA}.)


The argument is even clearer if a superconducting shield surrounds a solenoid. (See \cite{acr} for a qualitative analysis.) In this case, the flux of the solenoid must be quantized in multiples of $hc/2e$, hence $\delta \Lt$ is identically zero. Then we see that the only relevant coupling could be between  $\langle \Lt\rangle$ and $\Lo$, which is related to a nonlocal, instantaneous change of the velocity distribution as explained below in Sec. \ref{S3}.

In what follows, we display two distinct aspects of topological effects, one evolving continuously in time, the other instantaneous. The first aspect, which depends on the $(\delta \Lt)\Lo$ coupling, appears superficially to have a local description, while the instantaneous aspect, which depends on the $\langle \Lt\rangle \Lo$ coupling, has only a nonlocal description in terms of potentials.

\section{The continuous aspect}
\label{S2}

We first describe the continuous aspect of the Aharonov-Bohm effect. Consider an electron moving at constant velocity in a straight line, as it passes the uniformly charged cylinder of the previous section, rotating around the $z$ axis, along which runs a wire of uniform and opposite charge. (See Fig. 1.)  According to Ref. \cite{lev}, the electron's magnetic field induces a torque on the cylinder.  The torque, integrated over the angular displacements of the cylinder, yields a phase in the electron's wave function.  According to quantum mechanics, the phase arises from the vector potential of the cylinder, and implies no change in the velocity of the electron.  By contrast, Ref. \cite{lev} has no place for a vector potential, hence the change in the phase of the electron's wave function must imply a transient change in the electron's velocity---a change incompatible with quantum mechanics, which predicts a phase change with no corresponding physical effect.  (Likewise, in the simple model of the previous section, the approach of Ref. \cite{lev} would imply that one electron wave packet lags behind the other; but the lag would not necessarily be observable.) As a test of these incompatible predictions, we could prepare an ensemble of particles with a given mass and charge, and another ensemble of particles with equal mass but opposite charge, or no charge, in the initial state of Fig. 1.  After letting the ensembles evolve, we would measure the average position of the particles in each ensemble.  A transient difference in their velocities (due to the rotating cylinder) would induce a lag between these average positions.  Quantum mechanics predicts no such lag.

\begin{figure}
\centerline{
\includegraphics*[width=120mm]{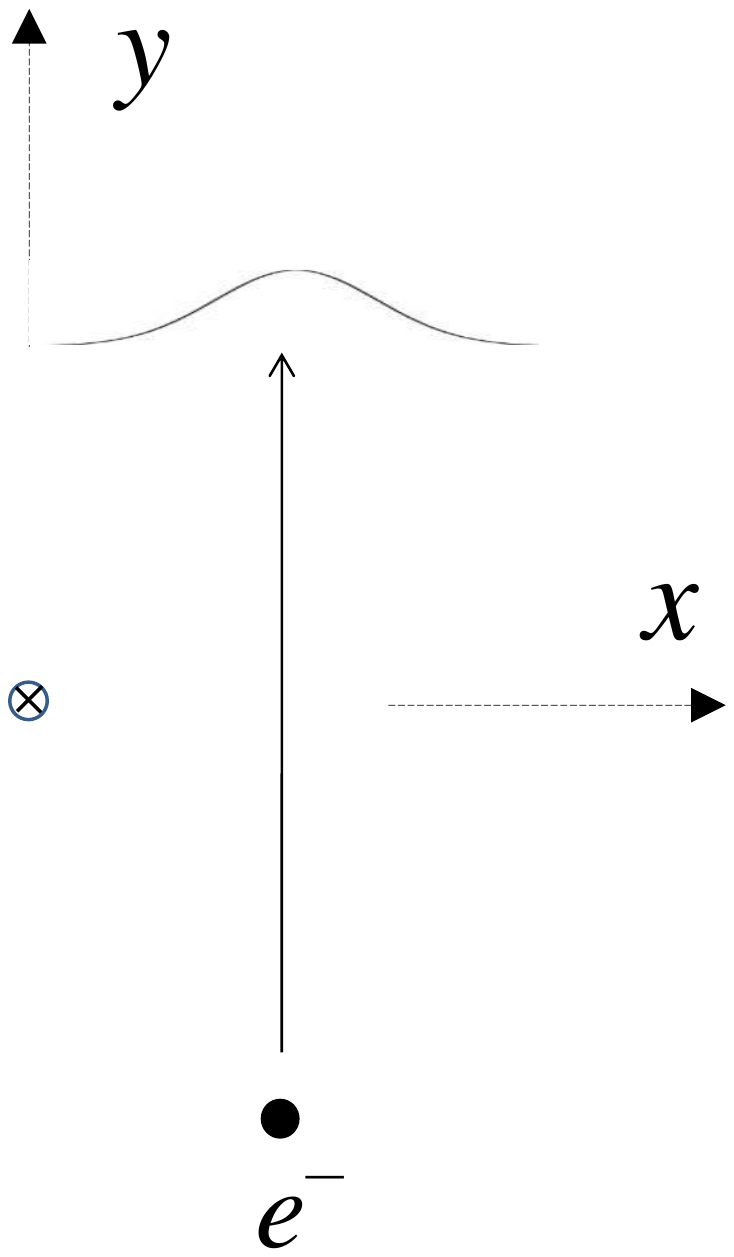}}
\caption[]{An electron passes a solenoid at the origin $x=0=y$ in a single wave packet.}
\label{Fig1}
\end{figure}

We now quantitatively analyze the dynamics of the electron and its entanglement. The flux of the cylinder depends on its angular momentum $\Lt$ and moment of inertia. A Hamiltonian for the electron is

\begin{equation}
H=\frac{(p_y-\mu \Lt/r)^2}{2m},
\end{equation}
where $p_y$ and $m$ are the electron's linear momentum and mass respectively, $r=\sqrt{x^2+y^2}$ and $\mu$ is a constant. Note $\Lt$ is unknown, and so is $p_y$, but
\begin{equation} \label{v}
v_y=\frac{p_y-\mu \Lt/r}{m}
\end{equation}
is a gauge-invariant constant of the motion.  In addition, we have
\begin{equation} \label{y}
{\dot \vt} = -{\mu}{{v_y}\over r} ~~~,
\end{equation}
where $\vt$, as defined above, is the canonical conjugate of $\Lt$.  Equations (\ref{v}-\ref{y}) define entanglement between the electron and flux. This entanglement changes continuously as a function of the distance between the electron and flux.

As proposed by Vaidman \cite{lev}, the entanglement between the electron and flux can be explained via the electromotive force exerted on the cylinder by the electron. However, this semi-classical analysis begs the question, why is the velocity of the electron in Fig. 1 constant (according to quantum mechanics)? (See \cite{acr} for related arguments.)

\section{The instantaneous aspect}
\label{S3}

We now describe the instantaneous aspect of the vector AB effect. Consider two interfering wave packets of an electron around an (inaccessible) solenoid, as in Fig. 2.  We recall the definition of a {\it modular variable} \cite{mod}.  Any continuous physical quantity can be expressed as a multiple of a constant (with the units of that physical quantity) plus a remainder, the $modular$ part of the quantity.  For example, if the physical quantity is displacement along the $x$-axis and the constant is $L$, then the modular displacement is $x$ mod $L$, which equals $x - nL$ for integer $n$ such that $0\le x-nL < L$.  Among modular variables, we focus on {\it modular velocity}.  Refs. \cite{ak,k} apply modular velocity to an analysis of the vector AB effect.  They show that the modular velocity of the diffracting electron changes abruptly at the moment when the electron wave packets and the enclosed flux fall on a single straight line.  In addition, Ref. \cite{k} offers two methods to measure the modular velocity of an electron diffracting around a solenoid between two slits in a screen.  This result changes our understanding of the AB effect; formerly, the vector AB effect was not associated with a precise time, because the effect was manifest only when the electron wave packets recombined.  Now, given this modular velocity effect \cite{ak} and the possibility of observing it, we can understand the AB phase as arising from an instantaneous interaction at a distance between the solenoid and the electron.  It is this action at a distance that ultimately rules out any explanation of the AB effect based on (local) forces.

We will demonstrate this effect in a convenient gauge.  In Fig. 2, an electron passes through an interferometer in two wave packets, separated by a distance $L$, with a solenoid located at the origin.  As the wave packets cross the $x$-axis on either side of the origin, in a line with the solenoid, their relative phase shifts by $e\Phi_B/\hbar c$ as a result of a (singular) vector potential $A_y$ that is nonzero only along the positive $x$-axis.  Defining the transverse modular velocity $v_x^{mod}$ as $v_x^{mod} = v_x ~{\rm mod}~ h/mL$, where $m$ is the electron mass, we have $mv_x^{mod} = p_x^{mod} \equiv p_x ~{\rm mod}~ h/L$; note that $A_x =0$ everywhere and at all times, hence $p_x$ is gauge invariant at all times.  We can calculate the expectation value of $v^{mod}_x$ by calculating the expectation value of $e^{ip_xL/\hbar}$ as a function of time; we will find that $p_x^{mod}$ changes abruptly as the wave packets cross the $x$-axis.

\begin{figure}
\centerline{
\includegraphics*[width=120mm]{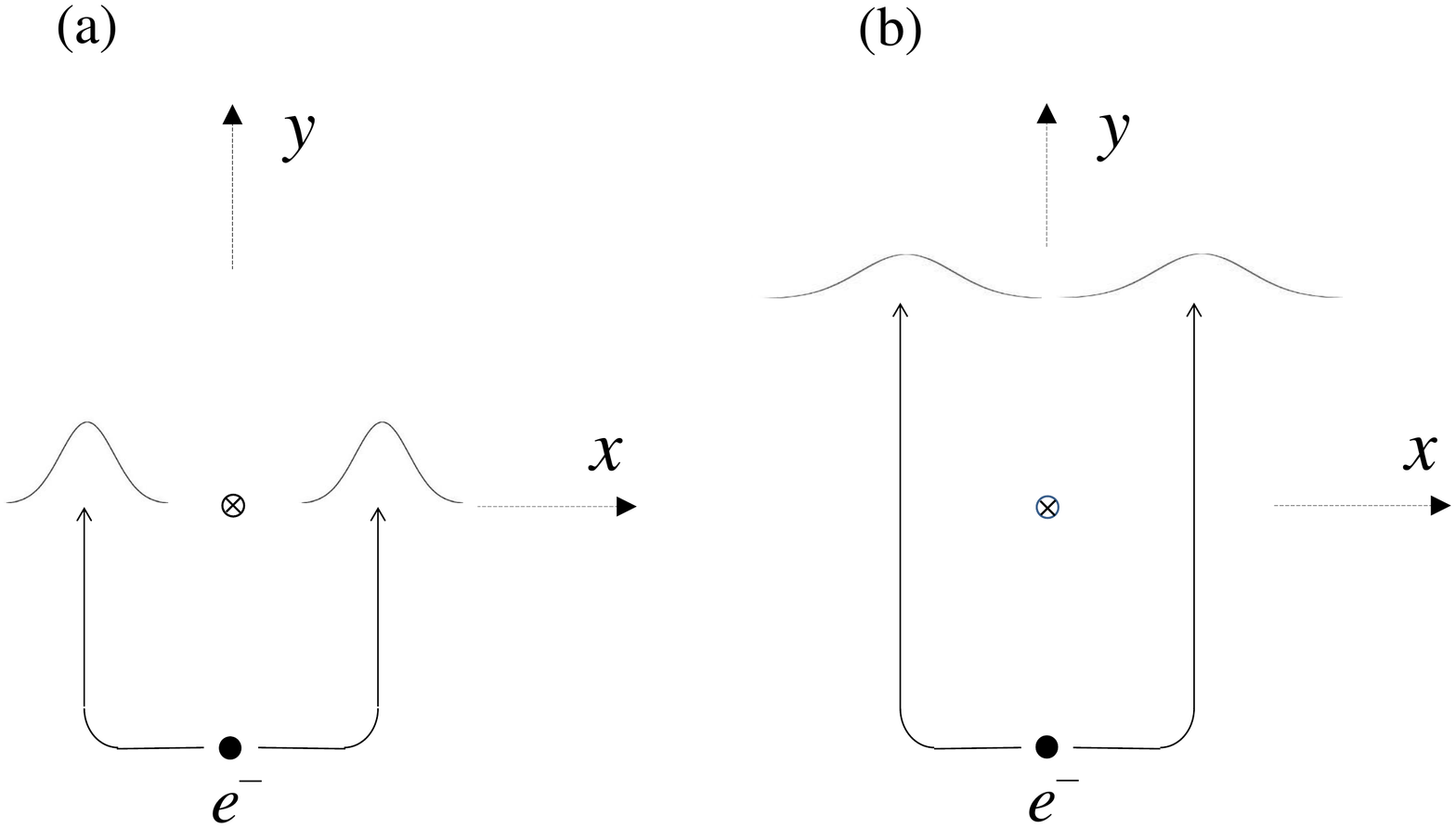}}
\caption[]{An electron passes through an interferometer, in two wave packets that enclose a solenoid at the origin $x=0=y$. The only vector potential is on the positive $x$-axis, where $A_y =\Phi_B \delta (y)\Theta (x)$, with the Heaviside function $\Theta (x) = 1/2 + x/2\vert x\vert$ and $d\Theta /dx =\delta (x)$.  (a) As the two wave packets cross the $x$-axis, the one on the right acquires a phase $e\Phi_B/\hbar c$ relative to the one on the left.\\ (b) The wave packets continue through the interferometer, spreading and ultimately overlapping and interfering; the interference pattern reveals the relative phase $e\Phi_B /\hbar c$.}
\label{Fig2}
\end{figure}

The wave packets evolve in three stages.  In the first stage, from their initial separation up to the approach to the $x$-axis, the Hamiltonian is simply $p^2/2m$ and the wave packets evolve freely with overall wave function $\Psi (x,y,t) = e^{-iHt/\hbar} \Psi (x,y,0)$.  Since $e^{ip_xL/\hbar}$ commutes with the Hamiltonian, its time-dependent expectation value, namely
\begin{equation}
\langle e^{ip_xL/\hbar} \rangle = \int dx~dy~
\Psi^* (x,y,0)e^{iHt/\hbar}
e^{ip_xL/\hbar}  e^{-iHt/\hbar} \Psi (x,y,0)
~~~,
\label{evo}
\end{equation}
cannot change during this stage; the modular transverse momentum cannot change.  In the second stage, as the wave packets briefly cross the $x$-axis, we can neglect their overlap, since the electron cannot enter the solenoid. (Because of the solenoid, the $xy$-plane is not simply connected.)  The wave packets have disjoint support as they cross the $x$-axis, and there is no difficulty in evaluating Eq. (\ref{evo}).  Each wave packet has its own Hamiltonian:  it is $p^2/2m$ on the left, while on the right it is
\begin{equation}
{{p_x^2}\over {2m}} +{1\over {2m}} \left[ p_y -{e\over c} A_y\right]^2
= {{p_x^2}\over {2m}} +{1\over {2m}} \left[ p_y -{e\over c} \Phi_B \delta (y)\right]^2~~~,
\end{equation}
which has eigenstates
\begin{equation}
e^{ik_x x} e^{ik_y y + ie\Phi_B\Theta (y)/\hbar c}~~~~.
\end{equation}
For $y>0$ the phase $e\Phi_B/\hbar c$ is independent of $k_x$ and $k_y$, hence it appears as an overall phase factor $e^{ie\Phi_B/\hbar c}$ multiplying the right wave packet.

In the third stage of the evolution, the Hamiltonian is again the free Hamiltonian, and again the expectation value of $e^{ip_xL/\hbar}$ cannot change, so it keeps whatever value it had after the wave packets crossed the $x$-axis.  The left and right wave packets continue to spread, overlap and interfere, revealing the AB effect.  The new insight that this analysis provides is that relative phase arises abruptly the moment the wave packets cross the $x$-axis, and that this sudden change in the phase offers a new way to observe the AB effect via a change of the velocity distribution \cite{ak}.  The change in modular momentum may not be immediately measurable.  Indeed, causality forbids instantaneous measurement of the change in phase.  Refs. \cite{ak,k,ar} discuss this causality constraint further.  What is essential here is that the sudden change in the distribution of modular momentum can, over time, be verified, ruling out any explanation of the effect as a local effect of forces.

An analysis via local forces cannot account for the behavior of velocities in these last two examples.  In the previous example, such an analysis implies that the velocity changes---and it does not; in the last example, the same analysis implies that the velocity does not change suddenly---and it does.  In both cases it is the vector potential that distinguishes momenta from velocities and determines the behavior of velocities.

\section{Conclusions}
\label{SC}
We have revisited the nonlocality of the Aharonov-Bohm effect. Proposed local descriptions \cite{lev, kk} prove incomplete when we consider e.g. an electron separated by superconducting shielding from a solenoid. We have shown that the Aharonov-Bohm effect has two distinct aspects, one continuous and one instantaneous. The latter is manifestly nonlocal; it underlines the necessity of describing quantum systems via gauge-dependent quantities rather than local forces, which cannot account for abrupt changes in modular velocity.

\begin{acknowledgments}

We thank Michael Berry for helpful comments.  Y.A. and E.C. thank the Israel Science Foundation (grant no. 1311/14) for support, and Y.A. acknowledges support also from the ICORE Excellence Center ``Circle of Light", and the German-Israeli Project Cooperation (DIP). E.C. acknowledges support from ERC AdG NLST. D.R. thanks the John Templeton Foundation (Project ID 43297) and the Israel Science Foundation (grant no. 1190/13) for support.

\end{acknowledgments}

\end{document}